\documentclass[prb,aps,twocolumn,superscriptaddress]{revtex4}

\usepackage{graphicx}
\usepackage{natbib}
\usepackage{mathrsfs}
\usepackage{amsmath}

\begin{document}

\title{High-pressure magnetization and NMR studies on $\alpha$-RuCl$_3$}

\author{Y. Cui}
\thanks{These authors contributed equally to this study.}
\affiliation{Department of Physics and Beijing Key Laboratory of
Opto-electronic Functional Materials $\&$ Micro-nano Devices, Renmin
University of China, Beijing, 100872, China}

\author{J. Zheng}
\thanks{These authors contributed equally to this study.}
\affiliation{Department of Physics and Beijing Key Laboratory of
Opto-electronic Functional Materials $\&$ Micro-nano Devices, Renmin
University of China, Beijing, 100872, China}
\affiliation{Department of Physics, Beijing Jiaotong University, Beijing
100044, China}

\author{K. Ran}
\affiliation{National Laboratory of Solid State Microstructures
and Department of Physics, Nanjing University, Nanjing 210093, China}

\author{Jinsheng Wen}
\affiliation{National Laboratory of Solid State Microstructures and Department
of Physics, Nanjing University, Nanjing 210093, China}
\affiliation{Innovative Center for Advanced Microstructures, Nanjing University,
Nanjing 210093, China}

\author{Zheng-Xin Liu}
\affiliation{Department of Physics and Beijing Key Laboratory of
Opto-electronic Functional Materials $\&$ Micro-nano Devices, Renmin
University of China, Beijing, 100872, China}

\author{B. Liu}
\affiliation{Department of Physics, Beijing Jiaotong University, Beijing
100044, China}

\author{Wenan Guo}
\affiliation{Department of Physics,Beijing Normal University,
Beijing, 100875, China}

\author{Weiqiang Yu}
\email{wqyu\_phy@ruc.edu.cn}
\affiliation{Department of Physics and Beijing Key Laboratory of
Opto-electronic Functional Materials $\&$ Micro-nano Devices, Renmin
University of China, Beijing, 100872, China}



\begin{abstract}

    We report high-pressure magnetization and $^{35}$Cl NMR studies on $\alpha$-RuCl$_3$
with pressure up to 1.5~GPa. At low pressures, the magnetic ordering
is identified by both the magnetization data and the NMR data, where
the $T_N$ shows a concave shape dependence with pressure. These data
suggest stacking rearrangement along the $c$-axis.
With increasing pressure, phase separation appears prominently at $P\ge$~0.45~GPa,
and the magnetic volume fraction is completely suppressed at $P\ge$~1.05~GPa.
Meanwhile, a phase-transition-like behavior emerges at high pressures in the remaining volume
by a sharp drop of magnetization $M(T)$ upon cooling, with the transition temperature $T_x$ increased to ~250~K at
1~GPa. The $1/^{35}T_1$ is reduced by over three orders of magnitude when cooled below 100~K. This characterizes
a high-pressure, low-temperature phase with nearly absent static susceptibility and
low-energy spin fluctuations. The nature of the high-pressure ground state is discussed,
where a magnetically disordered state is proposed as a candidate state.

\end{abstract}

\maketitle

\section{Introduction}

The Kitaev model in a honeycomb lattice, which is exactly solvable
with a spin-liquid ground state~\cite{Kitaev_ap_2006}, has
caused intensive research interests.
It was only recently discovered that the
iridates~\cite{Khaliullin_prl105_027204,Singh_PhysRevLett.108.127203,
Gretarsson_PhysRevLett.110.076402,Chun_np_2015,Kee_conmatphys_2016}
and the $\alpha$-RuCl$_3$~\cite{Pollini_PhysRevB.53.12769,
Plumb_PhysRevB.90.041112,Kee_PhysRevB.91.241110,Burch_Prl_2015,
Kindo_PhysRevB.91.094422,Sandilands_PhysRevB.93.075144,Koirzsch_PRL.117.126403,WenJS_prl_2016},
which have layered honeycomb lattice,
contain the Kitaev exchange interactions. Interplay of spin-orbital couplings,
Hubbard interactions, and Hund's coupling gives rise to an effective spin-1/2 quantum
compass model~\cite{Khaliullin_prl102_017205, Kim_science_2009,Foyevtsova_PhysRevB.88.035107,
Khaliullin_PhysRevB.94.064435,Haule_PhysRevLett.114.096403,LiJX_arXiv_2016}.
Although zigzag type magnetic ordering was found at low temperatures
in these systems because of the existence of non-Kitaev terms~\cite{Sears_PhysRevB.91.144420,Majumder_PhysRevB_91_180401,Johnson_PhysRevB_92_235119,Cao_PRB_93_13442},
proximate Kitaev spin liquid behaviors are manifested by fermionic and continuum-like excitations
observed in inelastic neutron and Raman scatterings~\cite{Banerjee_nm15_733,Burch_Prl_2015,Nasu_np_2016,Nagelar_2016_arxiv}.

\begin{figure}[t]
\includegraphics[width=6.5cm, height=7.5cm]{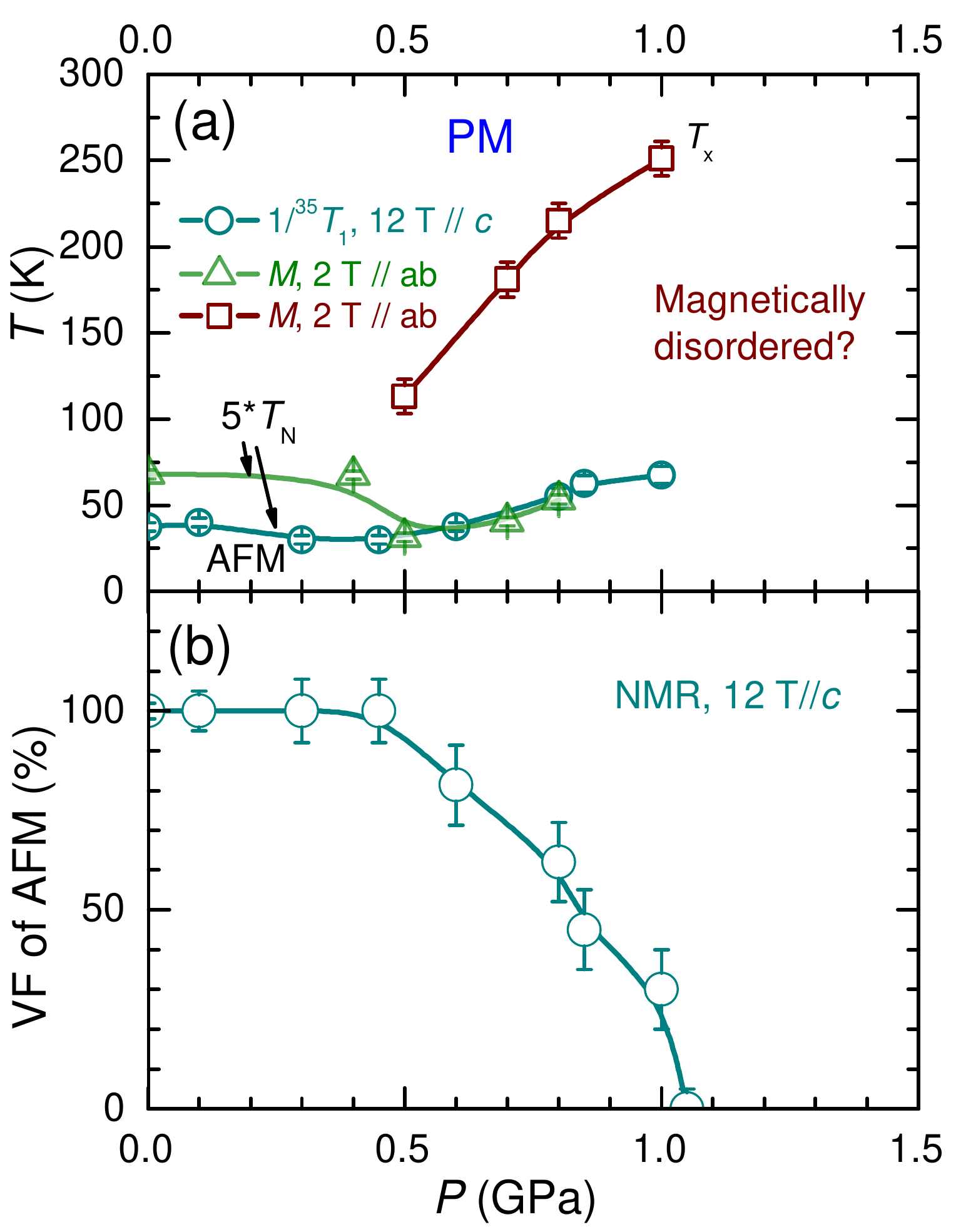}
\caption{\label{pd}(a) The ($P$, $T$) phase diagram of
$\alpha$-RuCl$_3$ determined by the $M(T)$ and the NMR data
under different field conditions.
$T_N$ represents the N\'{e}el transition to a
magnetically ordered phase (AFM) at low pressures,
and $T_x$ the high-temperature phase transition to
a magnetically disordered phase.
(b) The volume fraction (VF) of the AFM phase
as a function of pressure determined by the $^{35}$Cl
spectral weight at different phases.
}
\end{figure}

Since the ordered moment in the zigzag order is small~\cite{Banerjee_nm15_733},
spin-liquid-like behaviors may be observed if the magnetic
order is suppressed under various external conditions.
For instance, the magnetic
ordering in $\alpha$-RuCl$_3$ is suppressed by magnetic fields larger than 7.6~T
aligned in the $ab$-plane, where unconventional properties were reported
in the magnetically disordered
phase~\cite{Baek_PRL_119.037201,LeeM_PRL_118_2017,ZhengJC_arxiv_1703.08474,Kim_PRB_95_180411_2017,Corredor,Hess_arxiv_170308623}.
More surprisingly, a recent high-pressure specific heat measurement
on $\alpha$-RuCl$_3$ reveals that the N\'{e}el temperature $T_N$
is enhanced by pressure; above 0.7~GPa, however,
no signature of magnetic transition is seen at low temperatures,
where a spin-liquid ground state was proposed~\cite{SunLL_arxiv_170506139}.
High-temperature magnetic studies are highly desired to uncover
the true nature of this high-pressure phase.

In this paper, we report both the  magnetization $M(T)$
and the  $^{35}$Cl NMR studies on $\alpha$-RuCl$_3$
single crystals, with pressure up to 1.5~GPa. Our main findings are
summarized in the ($P$, $T$) phase diagram shown in Fig.~\ref{pd}(a).
At low pressures, a paramagnetic (PM) to a long-range-ordered
antiferromagnetic (AFM) phase is seen, but the magnetic transition
temperature $T_N$ first decreases with pressure, and
increases again when passing through 0.5~GPa. Meanwhile, the volume
fraction of this magnetic phase decreases with pressure, and is completely
suppressed at $P\ge$~1.05~GPa (Fig.~\ref{pd}(b)). Interestingly,
another phase-transition-like behavior is observed in the remaining sample volume,
with a very high onset temperature labelled as $T_x$.
$T_x$ is about 120~K at 0.5~GPa, and increases to 250~K at 1.0 GPa.
The emergent low-temperature, high-pressure phase has largely
reduced magnetization and  $1/^{35}T_1$, whose nature needs to be
further investigated.

The paper is organized as the following. Material growth and measurement
techniques are shown in Section~\ref{smm}. The magnetization
data are presented in Sec.~\ref{smh}. The NMR data with field applied along the $c$-axis
are presented in Sec.~\ref{snmrlp} for the low-pressure phase, and in Sec.~\ref{snmrhp}
for the high-pressure phase. Sec.~\ref{snmr3}  shows the
NMR measurements with field applied in the $ab$-plane.
Detailed discussions and possible scenarios for the high-pressure phase are given in Sec.~\ref{sd},
and a short summary is given in Sec.~\ref{summary}.

\section{Materials and methods \label{smm}}

The single crystals were grown by a chemical vapor transport
method~\cite{ZhengJC_arxiv_1703.08474}.
The high-pressure dc magnetization measurements were performed
in a SQUID (Quantum Design) with a 2 T field applied in the $ab$-plane of the
crystal, where a single crystal was placed in a BeCu pressure cell,
with daphne 7373 as the pressure medium.
For NMR measurements, a single crystal was placed in a NiCrAl pressure cell,
using the same pressure medium. The pressure values
presented in this article were calibrated by the Cu$_2$O NQR
frequency~\cite{Thompson_Cu2O_NQR_HP} measured at 5~K.

For NMR measurements, we primarily report the measurements with field applied along
the $c$-axis. The data with field applied in the $ab$-plane are also briefly
reported for comparison. The $^{35}$Cl (spin-3/2) spectra were taken by the standard spin-echo
sequence $\pi/2$$-$$\tau$$-$$\pi$ ($\pi/2$~$\approx$~1~$\mu$s).
The low-pressure phase has very short $^{35}T_1$ and short $^{35}T_2$ ($\approx$ 500 $\mu$s), and
the high-pressure phase has long $^{35}T_1$ and long $^{35}T_2$($\ge$~100~ms).
The low-pressure phase is selected by fast requisition after applying a $\pi/2$ pulse in advaced of the spin echo.
The high-pressure phase is selected by using very long $\tau$ ($\approx$~100~ms) in the spin echo sequence.
The spin-lattice relaxation rates are measured by the magnetization inversion-recovery method, with the
magnetization fit by $I(t)/I(\infty )=a$$-$$0.1e^{-(t/T_1)^\beta}$$-$$0.9e^{-(6t/T_1)^\beta}$.
Here $\beta\approx$~1 in the low-pressure paramagnetic phase, and ${\beta}\ge$~0.5
in the high-pressure phase. We tested that all our observations are reversible with pressure.

\section{Magnetization data and analysis \label{smh}}

\begin{figure}[t]
\includegraphics[width=8.5cm, height=7cm]{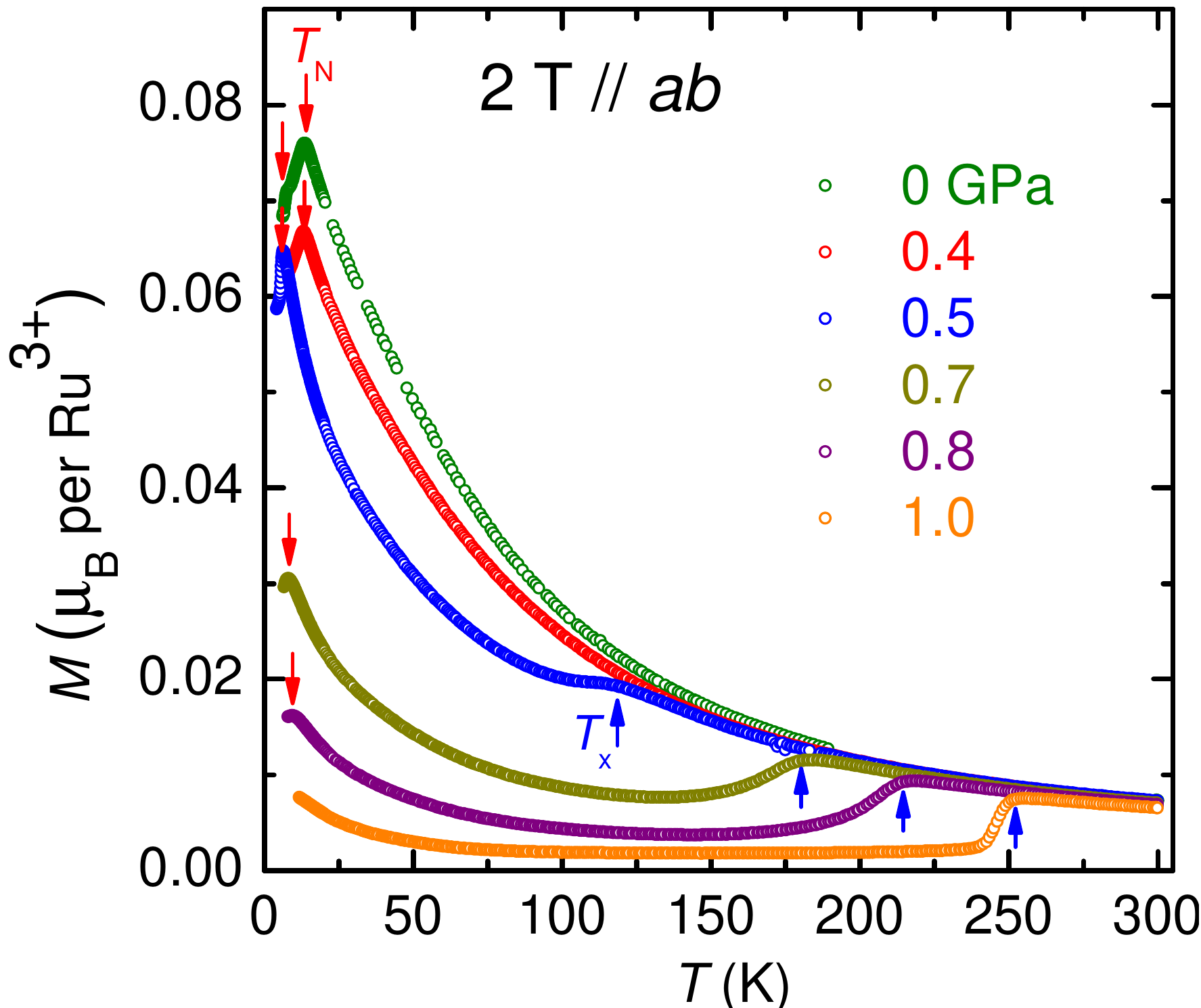}
\caption{\label{mh}The dc magnetization $M(T)$
measured as functions of temperatures at different pressures.
The field is set at 2~T in the $ab$-plane of the crystal.
The down arrows point at the $T_N$ by the peaked feature in $M(T)$,
and the up arrows point at an onset temperature $T_x$ for a sudden drop of $M$.
}
\end{figure}

The dc magnetization was measured in the pressure cell with a constant field of 2~T
applied in the $ab$-plane. The data are shown  in Fig.~\ref{mh}
with escalating pressures up to 1 GPa, the highest in the current system.
At zero pressure, the magnetic phase transitions
to the ordered states are identified by the peaked feature at $T_N\approx$13.6~K.
This transition temperature is consistent with $AB$ stacking
of the crystal~\cite{Banerjee_nm15_733}.
A second drop is also noticeable at 7.8~K, consistent with
the $ABC$ stacking~\cite{Banerjee_nm15_733}.

The transition temperature $T_N$, marked by the down arrows at positions of
sharp drops in $M(T)$, changes with pressure.
As shown in Fig.~\ref{pd}(a), $T_N$ first decreases with pressure, then increases weakly
with pressure at $P\ge$0.5 GPa, and becomes unresolvable at $P=$~1.0~GPa.

At pressure above 0.4~GPa, however, a second decrease in $M$ is seen when cooled below a high
temperature labelled as $T_x$. With increasing pressure, $T_x$ moves up quickly,
and the drop of $M$ below $T_x$ becomes more prominent.
$T_x$ climbs from 120~K at 0.5~GPa to 250~K at 1.0~GPa, and still does not saturate.
At 1.0 GPa, $M$ is reduced by 75$\%$ when cooled from 250~K to 240~K, which
indicates a new low-temperature, high-pressure phase with significantly reduced static susceptibility.
In particular, the rapid decrease of $M$ in such a narrow temperature window
strongly supports a phase transition at $T_x$.

\begin{figure*}[t]
\includegraphics[width=14cm, height=5.5cm]{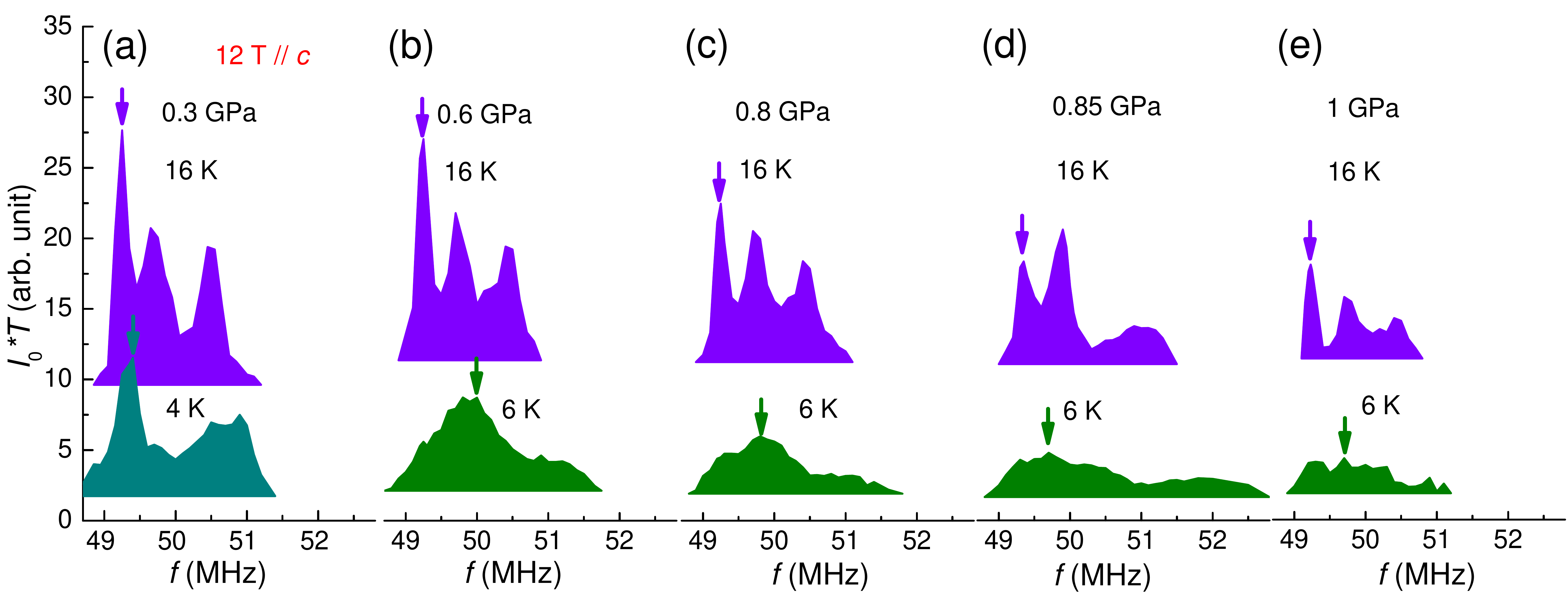}
\caption{\label{specf} (a)-(e): The $^{35}$Cl NMR spectra for
the AFM phase 16~K (above $T_N$) and 4~K/6~K (below $T_N$) at typical pressures.
All measurements are performed under 12~T field along the $c$-axis.
The same y-axis scale is used for all figures, with vertical offsets for clarity.
A $\tau$~$\approx$~35~$\mu$s is used in the echo sequence (Sec.~\ref{smm}).
The down arrows mark the position where the $^{35}T_1$ were measured and shown
in Fig.~\ref{invt1clp}.
}
\end{figure*}

At 0.4~GPa~${\le}P{\le}$~1.0~GPa, a phase separation below $T_x$ is suggested by
a double drop of $M$ at both $T_x$ and $T_N$, and a progressive reduction in $M$
with pressure below $T_x$. This is later revealed in our NMR study, where a long-$^{35}T_1$ phase is seen
at $P\ge$ 1.05 GPa, whereas both a long-$^{35}T_1$ phase and
a short-$^{35}T_1$ phase are resolved at 0.45~GPa~${\le}P{\le}$~1.05~GPa.

\section{$^{35}$Cl spectra and $1/^{35}T_1$ for the low-pressure antiferromagnetic phase\label{snmrlp} }

\begin{figure}[t]
\includegraphics[width=8cm, height=6.5cm]{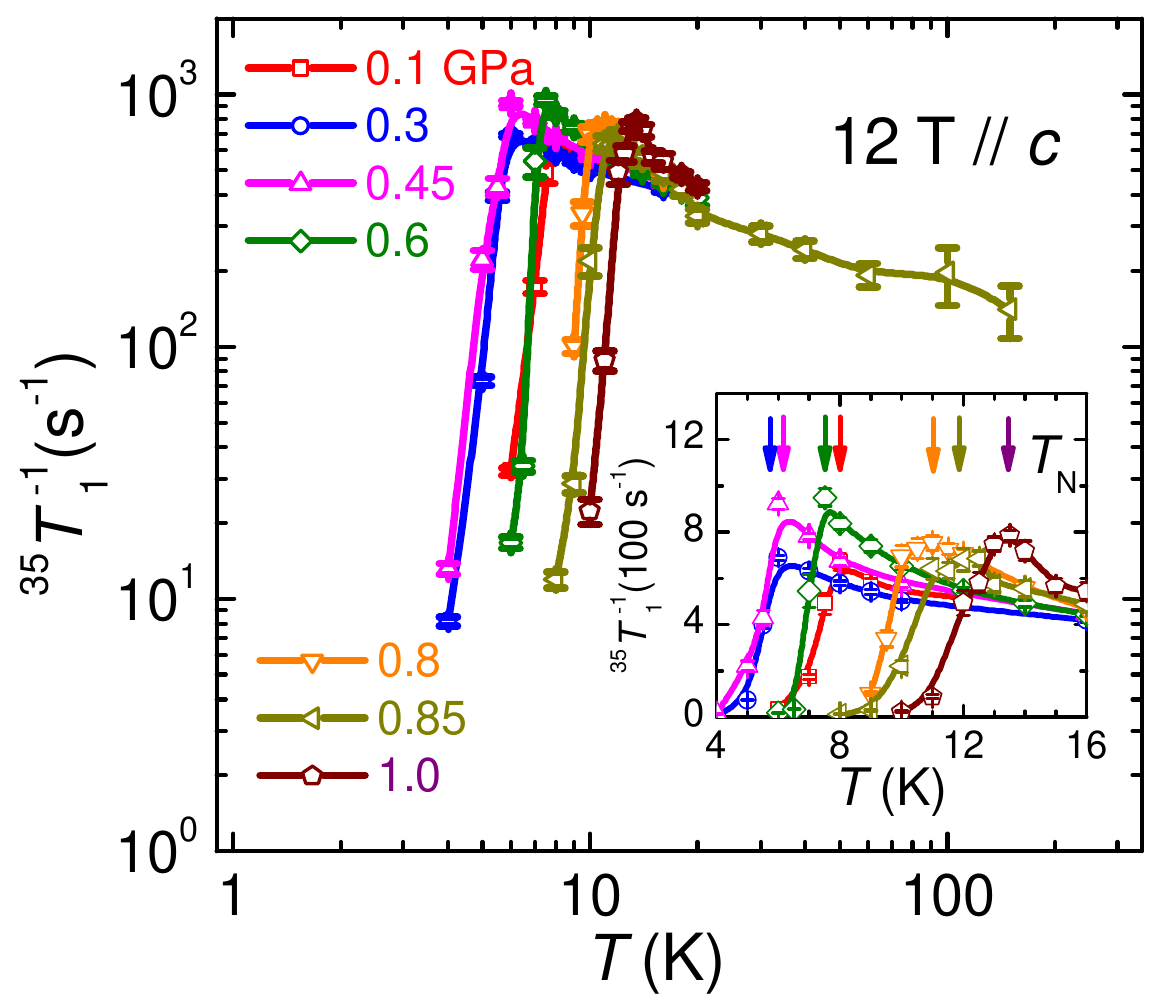}
\caption{\label{invt1clp} The spin-lattice relaxation
rate $1/^{35}T_1$ as functions of temperatures
for the low-pressure, short-$^{35}T_1$ phase.
The inset is an enlarged view of the low-temperature
data, with down arrows mark the $T_N$, characterized by the peaked feature in $1/^{35}T_1$ .
}
\end{figure}

\begin{figure}[t]
\includegraphics[width=7.5cm, height=6.5cm]{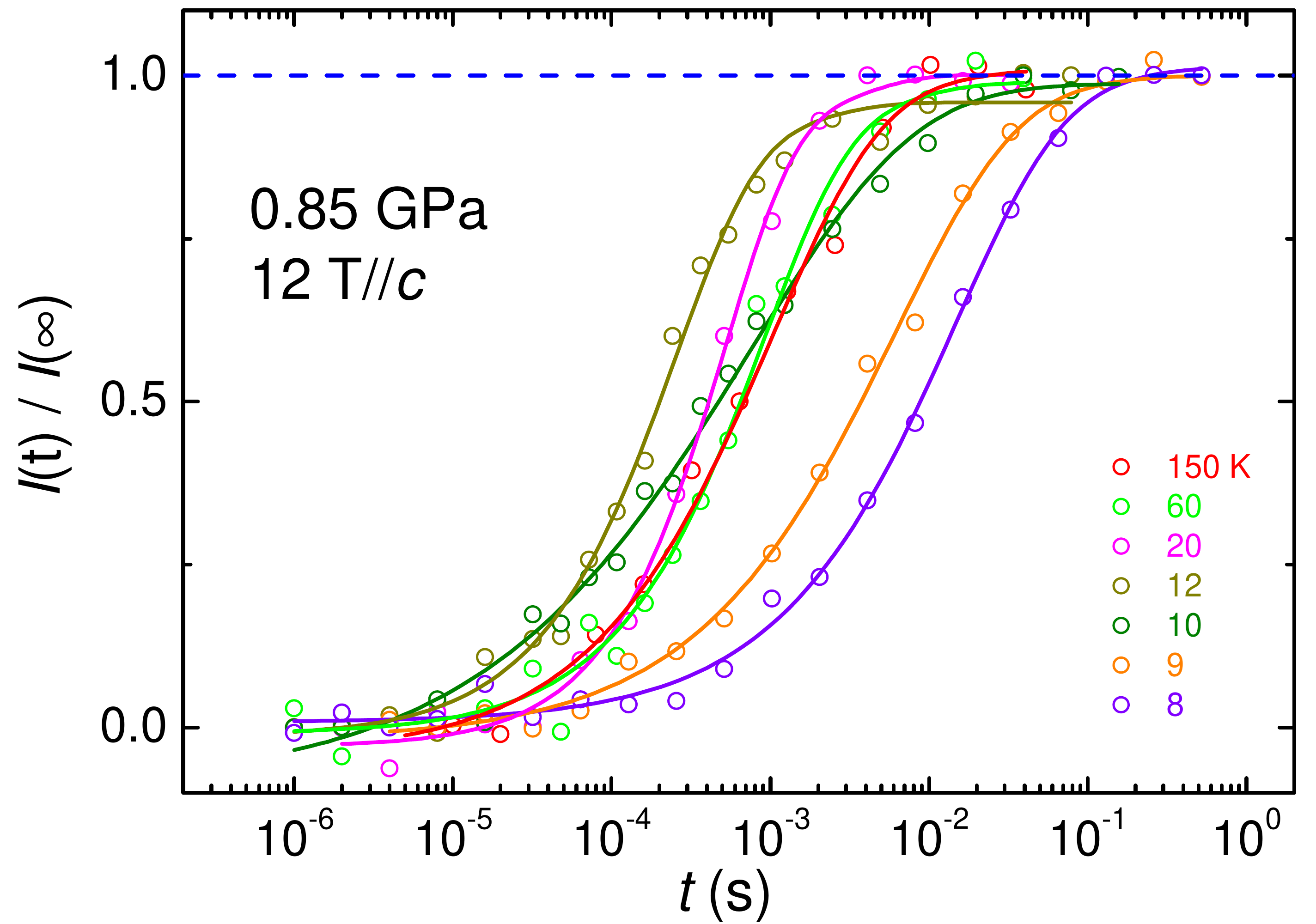}
\caption{\label{recoveryfast} Typical magnetization recovery as functions
of time for the $^{35}T_1$ measurements in the low-pressure phase.
The solid lines are the function fits described in Sec.~\ref{smm}.
}
\end{figure}

We first present data of a crystal measured under 12 T
field applied along the $c$-axis, in which case the $T_N$ is less
affected by field~\cite{Majumder_PhysRevB_91_180401,Johnson_PhysRevB_92_235119,Yadav_SciRep_2016}.
In the paramagnetic phase, nine NMR lines
are expected, given one center line and two satellites for each $^{35}$Cl site,
and three inequivalent $^{35}$Cl sites under magnetic field~\cite{ZhengJC_arxiv_1703.08474}.
As shown in Fig.~\ref{specf}(a), three center lines are
clearly observable at $T=$~16~K and $P=$~0.3~GPa, corresponding
to three $^{35}$Cl sites. The satellites are distant from the center lines (data not shown).
Their frequency shifts relative to $^{35}\gamma$$H$ are proportional
to $1/H$, due to a second-order quardrupolar correction.
Here $^{35}\gamma$$=$4.171~MHz/T is the gyromagnetic ratio
of $^{35}$Cl, and $H$ is the external field.
The NMR spectrum is broadened prominently when cooled to 4~K,
which is a clear signature of AFM ordering, due to the
development of the internal magnetic field.
Similar behaviors are also observed for other pressures
up to 1~GPa, with the broadened NMR spectra seen at
4~K/6~K (lower panels in Fig.~\ref{specf}(a)-(e)).

The $^{35}T_1$ are measured systematically on the leftmost peak marked by the down arrows in Fig.~\ref{specf}.
As shown in Fig.~\ref{invt1clp}, the $1/^{35}T_1$ are plotted as functions of temperatures
at different pressures. $1/^{35}T_1$ first increases upon cooling below 200~K, indicating
growing low-energy spin fluctuations toward magnetic ordering;
upon further cooling, it is peaked at the transition temperature
$T_N$ (see inset of Fig.~\ref{invt1clp}), followed by a dramatic decrease in the ordered state.
From this, the $T_N$ are precisely determined and plotted as a function of pressure in Fig.~\ref{pd}(a).
A non-monotonic change of $T_N$ is also revealed here, which will be compared with the
the magnetization data later.

The magnetization recovery curves and the high-quality of data fitting for $^{35}T_1$
are demonstrated in Fig.~\ref{recoveryfast}, under a typical pressure 0.85~GPa.
At 60~K, the magnetization already saturates with a recovery time of 0.02~s, which makes strong contrast
to the high-pressure phase presented later.

At 0.5~GPa~${\le}P{\le}$~1.0~GPa, the total spectral weight at 16 K for this
paramagnetic phase decreases with pressure, as shown in Fig.~\ref{specf}(a)-(e).
The spectral loss is not a $^{35}T_2$ effect, since a small $\tau$ (${\ll}T_2$) is used in the echo sequence (Sec.~\ref{smm}).
Similar spectral weight reduction is also seen at 4~K/6~K.
Such signal loss indicates a volume reduction in the measured phase.
The volume fraction of this phase is then calculated from the relative integrated spectral weight
at 16~K (same for 6~K), and plotted as a function of pressure in Fig.~\ref{pd}(b).
A drop of the volume fraction from 100$\%$ to zero is seen with pressure increased from 0.45~GPa to 1.05~GPa.

\section{$^{35}$Cl spectra and $1/^{35}T_1$ for the high-pressure, long-$^{35}T_1$ phase\label{snmrhp}}

\begin{figure}[t]
\includegraphics[width=7.5cm, height=6.5cm]{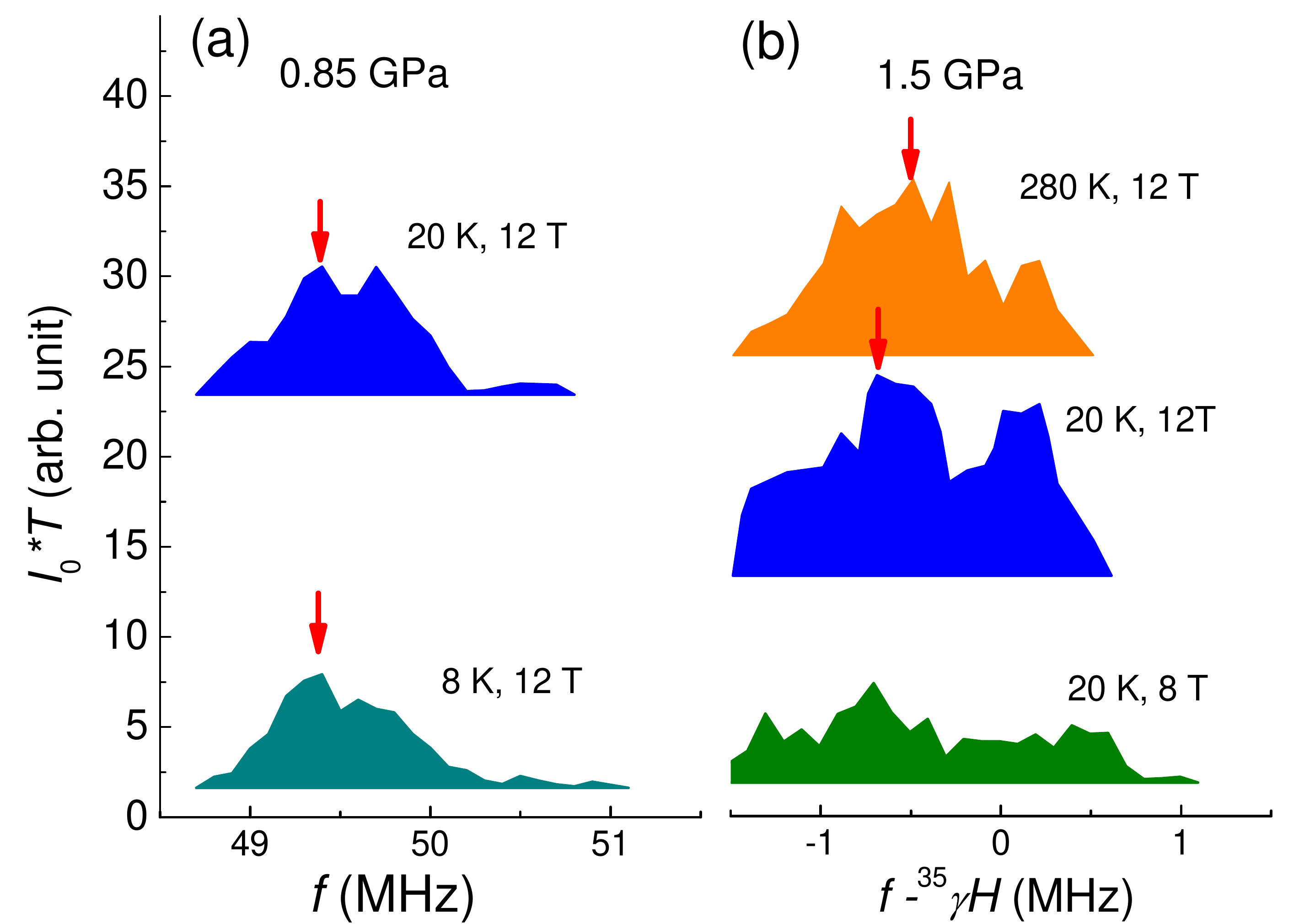}
\caption{\label{specs}
The $^{35}$Cl spectra of the long $^{35}T_1$ phase measured
at (a) 0.85~GPa and (b) 1.5~GPa, under different fields.
A $\tau\approx$~100~ms is used in the echo sequence (Sec.~\ref{smm}).
Vertical offsets are applied for clarity. The down arrows mark
the selected frequency for the $^{35}T_1$ measurements reported in Fig.~\ref{invt1chp}.
}
\end{figure}

\begin{figure}[t]
\includegraphics[width=7.5cm, height=6.5cm]{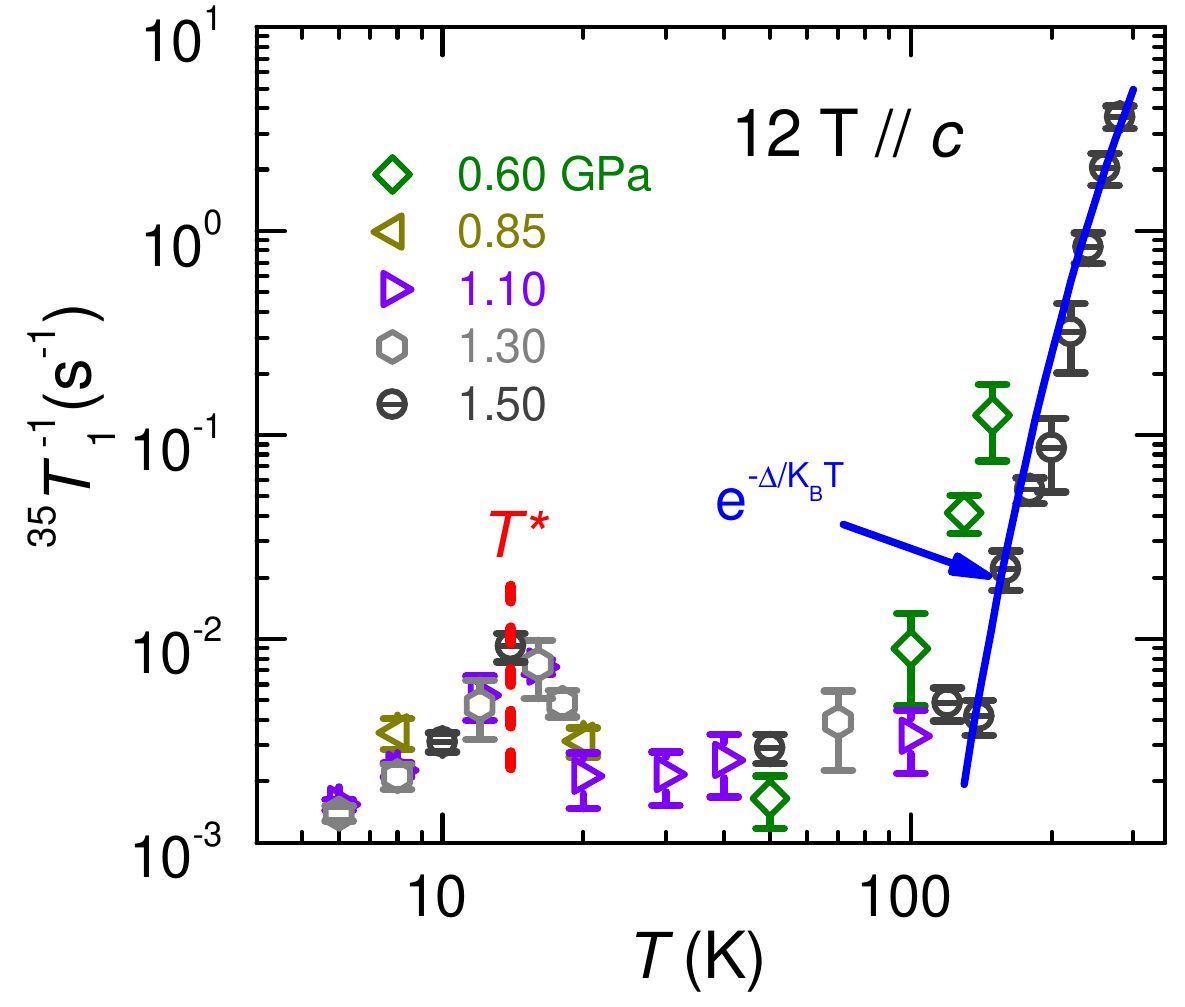}
\caption{\label{invt1chp} The spin-lattice relaxation
rate $1/^{35}T_1$ as functions of temperatures
for the long-$^{35}T_1$ phase.
The solid line is a function fit with a thermal activation form $1/T_1(T)=be^{-\Delta/K_BT}$,
with temperature from 280~K to 140~K.
The $T^*$ marks a temperature with a peaked feature in $1/^{35}T_1$.
}
\end{figure}

\begin{figure}[t]
\includegraphics[width=7.5cm, height=6cm]{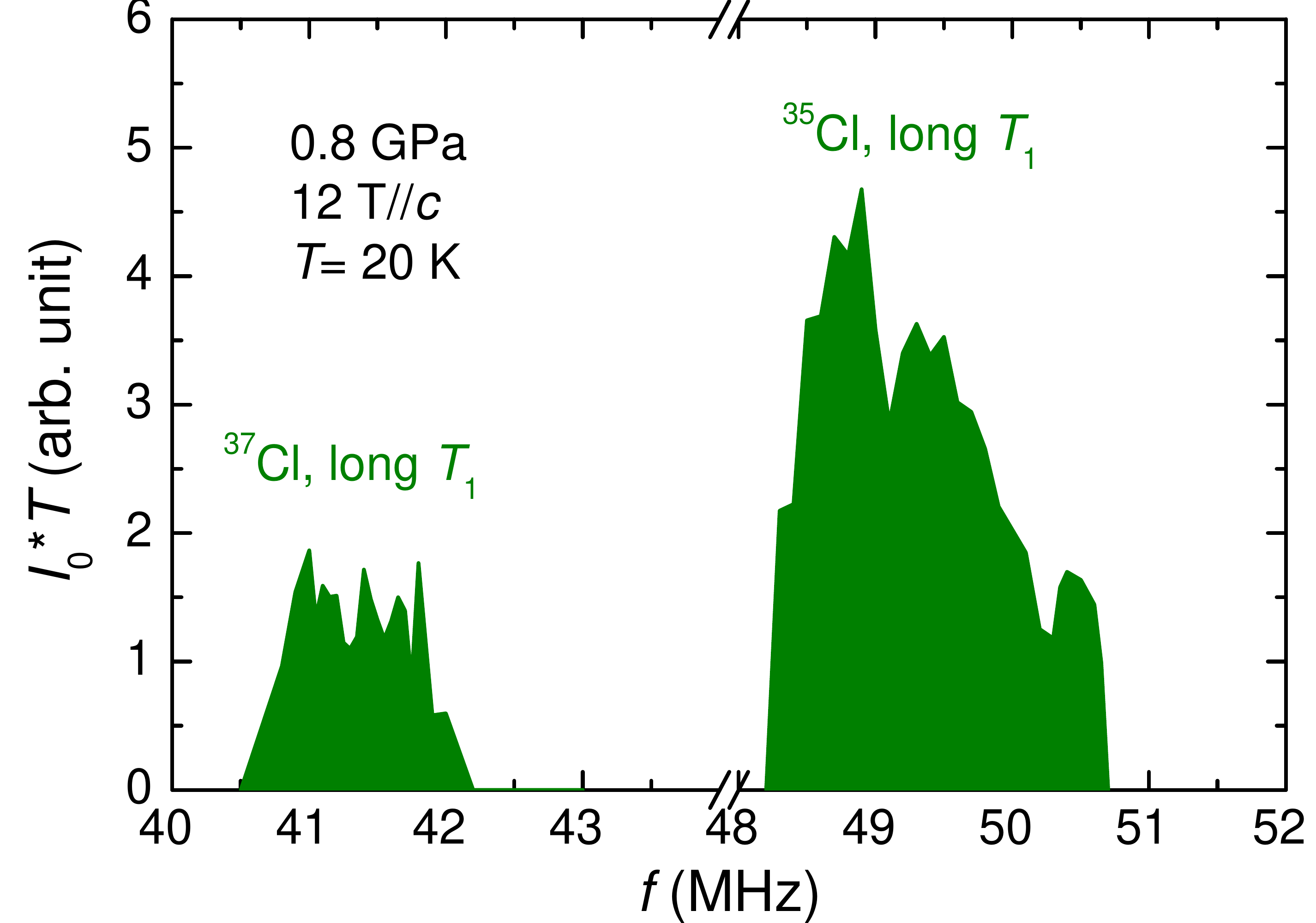}
\caption{\label{cl3537} The $^{35}$Cl and $^{37}$Cl spectra for the long-$T_1$ phase measured at 0.8 GPa.
}
\end{figure}

\begin{figure}[t]
\includegraphics[width=7.5cm, height=6cm]{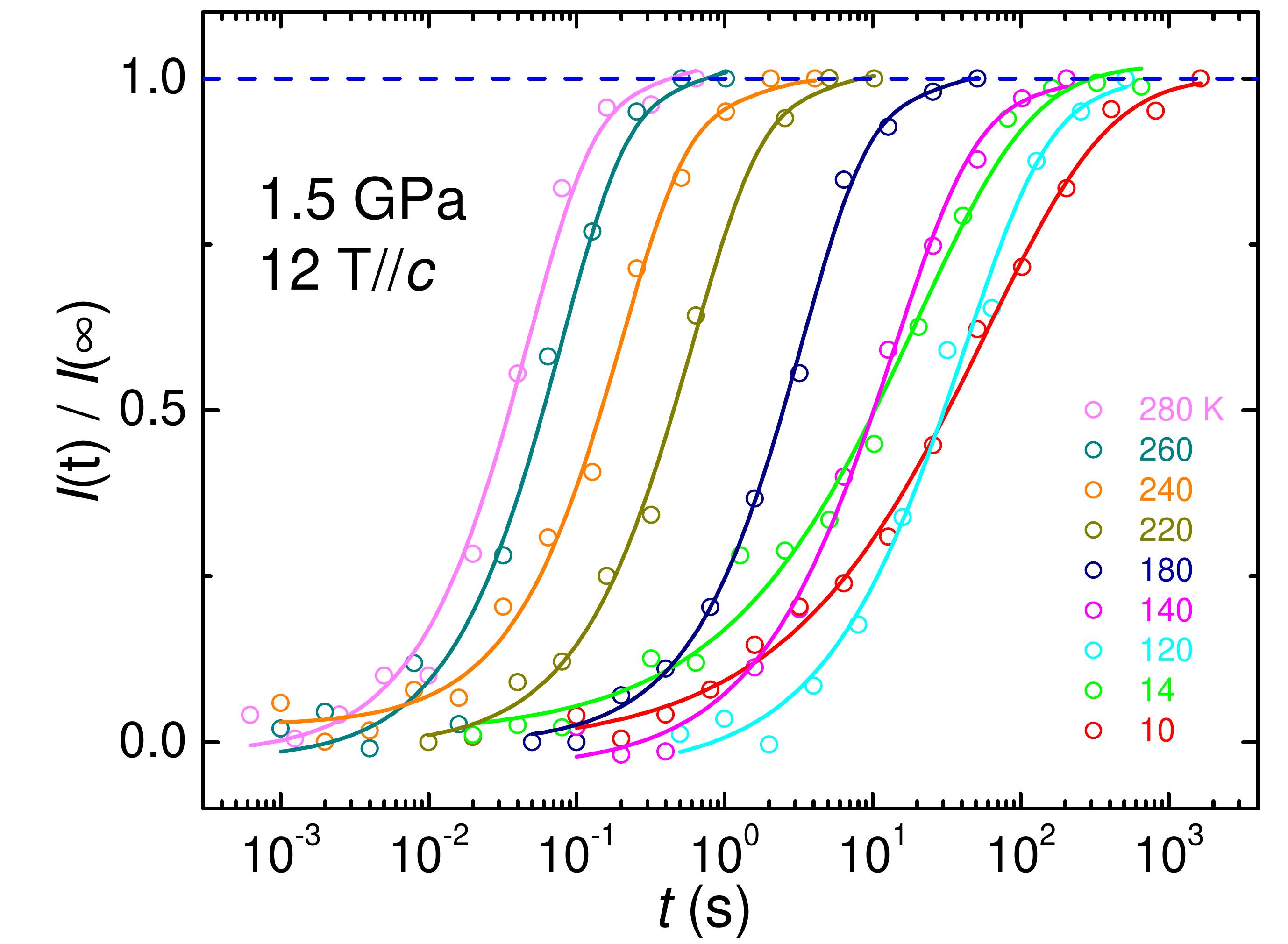}
\caption{\label{recoveryslow} The magnetization recovery curves for the $^{35}T_1$ measurements performed
at 1.5 GPa. The solid lines are function fits described in Sec.~\ref{smm}.
}
\end{figure}

\begin{figure}[t]
\includegraphics[width=7.5cm, height=6cm]{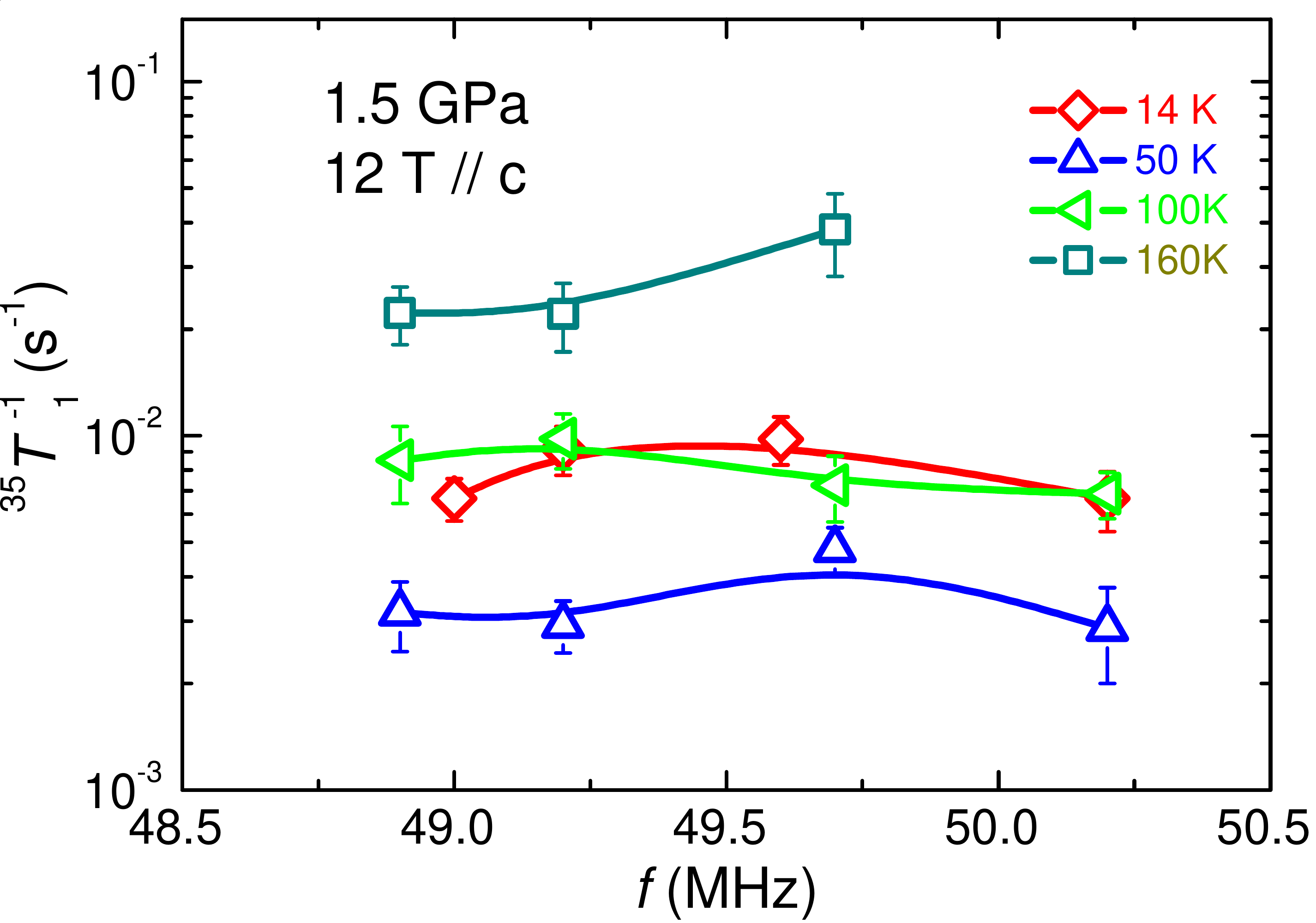}
\caption{\label{invt1vsf} The frequency dependence of $1/^{35}T_1$ across the NMR spectra,
measured at different temperatures under a representative pressure 1.5~GPa.
}
\end{figure}

The above reduced spectral weight is in fact transferred to an emergent
high-pressure phase with very long $T_1$ and $T_2$.
Taking advantage of their different $^{35}T_1$ and $^{35}T_2$,
the spectra of two phases are easily separable (see Sec.~\ref{smm}). In Fig.~\ref{specs}(a) and (b), the
long $^{35}T_1$ spectra are shown at 0.85~GPa and 1.5~GPa respectively.
For $P=$1.5~GPa, only the long-$^{35}T_1$ component is detectable by NMR.
For $P=$0.85~GPa, both a short-$^{35}T_1$ component (see Fig.~\ref{specf}(d)) and a long-$^{35}T_1$
component (see Fig.~\ref{specs}(a)) are resolved.
Here, the $30\%$ spectral weight loss in the short-$^{35}T_1$ phase
is exactly recovered in the long-$^{35}T_1$ phase.
Therefore, using NMR as a local probe, the phase separation of two phases
at intermediate pressures is clearly demonstrated.

The $1/^{35}T_1$ for the high-pressure phase are shown in Fig.~\ref{invt1chp}, as functions of temperatures.
For all pressures, the $1/^{35}T_1$ first decreases rapidly upon cooling,
and then levels off below 100~K.
Strikingly, below 100~K, the $^{35}T_1$ of the high-pressure phase is lowered by
about five orders of magnitude, compared to the low-pressure data presented in Fig.~\ref{invt1clp}.

In particular at 1.5~GPa, the $1/^{35}T_1$ is reduced by almost three orders of
magnitude when cooled from 280~K to 140~K, which suggests a phase transition
occurring at a temperature above 280~K.
We attempted to fit the data with a thermal activation function $1/T_1=be^{-\Delta/K_BT}$.
If a constant spin gap is considered,  $\Delta=150{\pm10}$~mev is estimated (see
the blue line in Fig.~\ref{invt1chp}). However, this gap value may be largely overestimated
in such a narrow high-temperature range.

For other pressures, a significant reduction in $1/^{35}T_1$ is also seen,
except that the reduction shifts to lower temperatures with decreasing pressure,
as shown by the data at 0.6~GPa (Fig.~\ref{invt1chp}).
These $1/^{35}T_1$ data are consistent with the magnetization
data, where a $T_x$ found to be 250 K at 1.0 GPa and keeps increasing with pressure.
By NMR, we are unable to trace the transition temperature, because of the poor signal quality at high temperatures.

To verify that our assignment of the $^{35}$Cl spectra is correct, the center lines of both $^{35}$Cl and $^{37}$Cl
isotopes are shown in Fig.~\ref{cl3537} for the long-$T_1$ phase, at a typical temperature (20~K) and a
typical pressure (0.8~GPa). Their center frequencies scale approximately with their gyromagnetic
ratios, $^{35}\gamma$=4.171 and $^{37}\gamma$=3.472.
The spectra weight for $^{35}$Cl is much larger, which is also consistent with
their different natural abundance (ratio $\approx$ 3:1).

In Fig.~\ref{recoveryslow}, typical magnetization recovery curves for the $^{35}T_1$ measurements
in the high-pressure phase are also shown. Below 140 K, the magnetization only becomes
appreciable when the recovery time is much longer than 1~s, in strong contrast to the low-pressure phase presented earlier.
The data fitting, as shown by the solid lines, is again of high quality.

The $1/^{35}T_1$ is checked as functions of frequency across the spectra, with data
shown in Fig.~\ref{invt1vsf} at a typical pressure 1.5~GPa. For any temperature,
the variation of $1/^{35}T_1$ across the whole spectra is less than a factor of two.
Therefore, the temperature behavior of $1/^{35}T_1$, presented in Fig.~\ref{invt1chp},
is validated.

\section{NMR spectra and $1/^{35}T_1$ data with field applied in the $ab$-plane \label{snmr3}}

\begin{figure}[t]
\includegraphics[width=8.5cm, height=7cm]{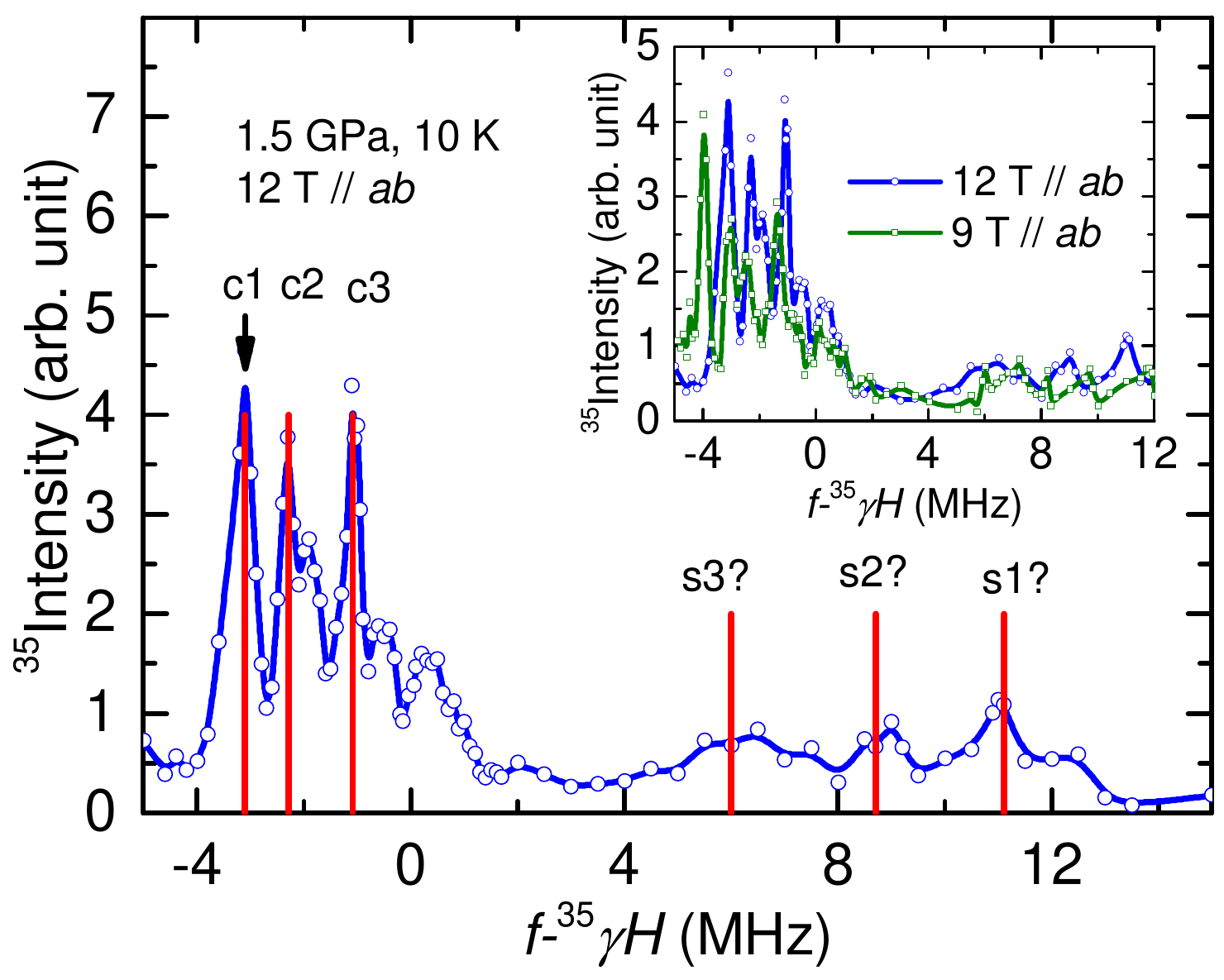}
\caption{\label{specab}
The $^{35}$Cl spectrum measured under 12 T field applied in the $ab$-plane,
at $P=$~1.5~GPa and at $T=$~10~K. The solid vertical lines mark the position
of center lines (c1,c2, and c3) and possible satellites (s1,s2, and s3) of
three inequivalent $^{35}$Cl sites under field.
The down arrow marks the position where the $1/^{35}T_1$ are reported in Fig.~\ref{invt1ab}.
Inset: The $^{35}$Cl spectra measured under 12~T and 9~T field at the
same pressure and temperature as shown in the main panel.
}
\end{figure}

\begin{figure}[t]
\includegraphics[width=7.5cm, height=6.5cm]{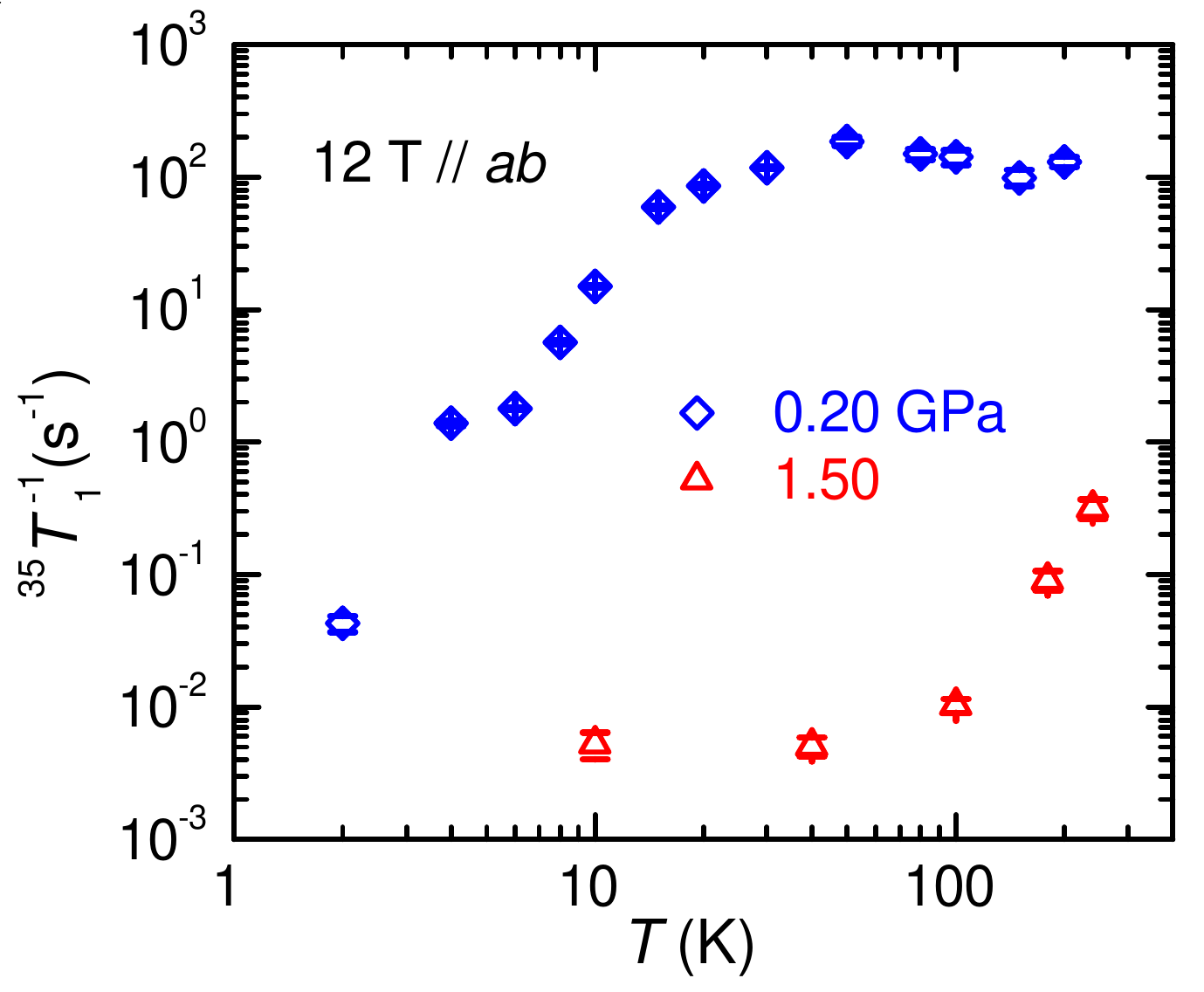}
\caption{\label{invt1ab}
 The spin-lattice relaxation rate $1/^{35}T_1$ as functions of temperatures
for the low-pressure (short-$T_1$) phase and the high-pressure (long-$T_1$) phase, with field applied in the $ab$-plane.
}
\end{figure}

For verification, we further performed NMR measurements with field applied in the $ab$-plane.
In Fig.~\ref{specab}, the $^{35}$Cl spectrum at 12~T is shown, under a fixed pressure
of 1.5~GPa and a fixed temperature 10~K.

Three sharp peaks are resolved close to the frequency $f_0$$=$$^{35}$${\gamma}H$,
and ranged from -4~MHz to 0~MHz. We confidently assigned them to the center lines of three
inequivalent $^{35}$Cl sites in the lattice, as labeled by c1, c2, and c3.
This is verified by comparing the spectra at 9 T and 12 T, as shown in the inset of
Fig.~\ref{specab}. For each peak, its relative frequency to $f_0$  scales with $1/H$,
fully consistent with a second-order quadrupolar correction in the absence of Knight shift~\cite{Abragam}.

We attempt to assign three broad lines at higher frequencies, labeled as s1, s2, s3,
to the satellite lines of three $^{35}$Cl sites. By comparing the spectra at 9 T and 12 T (inset of Fig.~\ref{specab}),
the relative frequency of each peak is consistent with a combination of a first-order (a large constant term)
and a second-order (a small term $\propto$$1/H$) quadrupolar correction~\cite{Abragam}.
However, the integrated spectral weight for s1, s2, s3 are much less than that of c1, c2, and c3.
This may be caused by local disorders, where the satellite signals are distributed broadly into backgrounds.
We defer this for future verification.

The $1/^{35}T_1$ with the in-plane field are measured at two typical pressures 0.2~GPa
and 1.5~GPa, with data shown in Fig.~\ref{invt1ab}. At 0.2~GPa, only a short-$^{35}T_1$ phase
is seen. The lack of a peaked feature in $1/^{35}T_1$ is consistent with the suppression of
the magnetic ordering by large in-plane fields~\cite{LeeM_PRL_118_2017,ZhengJC_arxiv_1703.08474,Kim_PRB_95_180411_2017,Corredor,Hess_arxiv_170308623}.
At 1.5 GPa, a sharp drop of $1/^{35}T_1$ is seen below 240 K, consistent with
the long-$^{35}T_1$ phase revealed with the $c$-axis field (Fig.~\ref{invt1chp}).

\section{Discussions \label{sd}}

The lines of phase transitions by different measurements are summarized in Fig.~\ref{pd}(a).
In the following, we first discuss the low-pressure, short-$^{35}T_1$ phase,
and then discuss the emergent high-pressure, long-$^{35}T_1$ phase.

Our NMR measurements under a $c$-axis field resolve a decrease of $T_N$ from 7.5~K to 6~K
with pressure from 0 to 0.45 GPa, and then an increase from 6~K to 14~K with pressure from 0.45~GPa to 1.0~GPa (see Fig.~\ref{pd}(a)).
Since the $c$-axis field barely affects the magnetic ordering\cite{Majumder_PhysRevB_91_180401}, the $T_N$ line determined
by NMR with $c$-axis field should be close to that at zero field.
This non-monotonic change of $T_N$ cannot be understood simply by an increased interlayer coupling.
In fact, $T_N$ varies with different c-axis staking pattern at the ambient pressure, with
$T_N$$\approx$~7.5~K for the ABC staking, and $T_N$$\approx$~14~K for the AB stacking~\cite{Banerjee_nm15_733}.
Coincidentally, our $T_N$ is about 7.5~K at $P$$=$~0, and levels off at about 14~K at 0.8~GPa~$\le$$P$$\le$~1.0~GPa.
This suggests that pressure causes an interlayer sliding, which transforms the ABC stacking to the AB stacking
under pressure. The dip of $T_N$ at 0.5~GPa is attributed an intermediate stacking, and/or
stacking fault in the middle. In fact, it was reported that $\alpha$-RuCl$_3$
is subject to the ABC-to-AB stacking rearrangement under mechanical pressure~\cite{Cao_PRB_93_13442}.

Although our crystal is verified to be primarily of $ABC$ stacking
after growth~\cite{WenJS_prl_2016,ZhengJC_arxiv_1703.08474},
the high-pressure magnetization measurement gives different $T_N$s with pressure below 0.5 GPa.
A primary $T_N$ of 13.6~K is obtained with sample sealed in the pressure cell,
which is consistent with the $AB$ stacking. The majority of the $AB$
stacking revealed in the $M(T)$ data suggests a strong impact of sample environment in the pressure cell.
In fact, the sample space in the magnetization pressure cell has a dimension of $\Phi$$2$$\times$4 mm, much
smaller than that of the NMR pressure cell ($\Phi$$4$$\times$8 mm).
For comparison, the $T_N$s measured by the the heat capacity measurements
in the diamond anvil cell~\cite{SunLL_arxiv_170506139} are much smaller than ours.
These facts suggest that the stacking rearrangement or stacking faults are more severe for
small cells.

In principle, the volume fraction of the low-pressure, antiferromagnetic phase can also be estimated
by the the $M(T)$ data by a two-component fitting scheme.
However, we found this fitting gives very large errors (data not shown), since only a narrow
temperature range is available to fit the low-temperature $M(T)$ data.
Nevertheless, at 1.0 GPa, the reduction of $M$ by 75$\%$ from 250 K to 240 K
is consistent with 30$\%$ volume fraction of the low-pressure phase estimated by NMR.

Next, we discuss the high-pressure phase. The high-pressure, low-temperature phase
is characterized by the strong reduction in both $M$ and the $1/T_1$,
which indicates that both the static susceptibility and the low-energy spin fluctuations
are strongly suppressed.
The phase transition is suggested by a rapid suppression of $M$ below $T_x$
at 1.0~GPa, and by a rapid drop of $1/^{35}T_1$ by about three orders of magnitude
when cooled from 280~K to 140~K at 1.5~GPa. The transition, if exist, also moves to higher temperatures
with pressure from both the $M$ and the $1/^{35}T_1$ data.
In the following, we discuss candidate mechanism, in our knowledge, to understand
the high-pressure phase.

i) {\it Transition to a nonmagnetic valence state of Ru ions}. To our knowledge, no such valence
transition has been reported for Ru$^{3+}$, in particular by such low pressures ($\sim$0.5 GPa).
However, it is still worthwhile to check if other valence states
are formed by covalent bonding or by charge ordering in $\alpha$-RuCl$_3$.

ii){\it Crossover from a low-spin to a high-spin state}. Since the transition-like
behavior at 250 K ($P$=~1.0 GPa) suggests a large pressure-enhanced energy scale,
it is important to know if a high-spin state is achieved under pressure.
However, Ru$^{3+}$, with $3d^5$ electronic state, has a low spin state ($s$=1/2) at zero pressure~\cite{Plumb_PhysRevB.90.041112}.
To our knowledge, a low-spin to a high-spin transition occurs with enlarged lattice parameters, in opposite to
the pressure effect.

iii){\it Formation of a gapped spin liquid}. A gapped spin liquid causes a large drop in $1/^{35}T_1$ and $M$.
However, the onset temperature $T_x$ are too high for a 2D spin liquid. Further,
the spin liquid usually occurs by a crossover behavior, rather than any phase-transition-like behavior
revealed in $M$ at 1.0~GPa.

iv){\it A structural Phase transition or strong structural disorder.}
We comment further on the interplay of the lattice structure and the magnetism in the high-pressure phase.
From the Kramers theorem~\cite{kramers}, the spin of Ru$^{3+}$ in the $4d^5$ state should remain finite, regardless
of local structures under pressure. Furthermore, the significant reduction on the static susceptibility
at high temperatures ($T_x$) is not consistent with a conventional lattice disorder, because
$M$ is usually enhanced by disorder in the antiferromagnetic materials.
Therefore, the sharp phase-transition-like
behavior in $M$ has to be understood with strong magnetic interactions.
On the other hand, the large $T_x$ (250~K at 1.0~GPa) suggests that the exchange coupling has an energy scale over 21.7~meV.
Since this is much bigger than any exchange couplings at the ambient pressure~\cite{Banerjee_nm15_733, WenJS_prl_2016},
a large change of local lattice is required to understand the data.

For $\alpha$-RuCl$_3$, pressure may affect its lattice structure in multifold ways.
First, pressure enhances the interlayer exchange couplings significantly.
Second, stacking faults can be induced by pressure. Even at the ambient pressure, the stacking pattern
causes a monoclinic to rhombohedral structural phase transition upon cooling, and
a coexistence of two structures between 70 K and 170 K~\cite{Glamazda_prb_2017}.
Our low-pressure evidence of different $T_N$ under different measurements, and the non-monotonic change of $T_N$
with pressure, suggest that the $c$-axis stacking is very sensitive to pressure conditions. At high pressures,
stacking fault is highly possible and induces inhomogeneous exchange couplings.
Lastly, as we will discuss later, local lattice dimerization in the $ab$-plane is also possible under pressure.
Although high-pressure x-ray measurements did not reveal any structural
transitional at 300~K~\cite{SunLL_arxiv_170506139}, low-temperature structural investigations are demanded to
reveal the primary structural cause for the dramatic increase of local exchange couplings.

v){\it Phase transition to a magnetically ordered state}.
NMR line broadening or splitting should be seen if magnetic ordering exists.
However, this is not observed in our system with both field orientations.
In particular with $H$$\parallel$$ab$, three sharp NMR lines are resolve at 1.5~GPa (fig.~\ref{specab}),
whose resonance frequencies are fully consistent with second-order quadrupolar corrections,
as described earlier by comparing with data under different fields. The absence of hyperfine field
is consistent with reduced local susceptibility revealed by the magnetization measurements.
The same conclusion is also drawn with field applied along the $c$-axis.
As shown in Fig.~\ref{specs}(b), the NMR linewidth at 1.5~GPa and 20 K,
measured at 8~T and 12~T, approximately scales with $1/H$ (see Fig.~\ref{specs}(b)),
again consistent with a second-order quadrupolar correction to the spectra.

However, in this high-pressure phase, a peaked behavior in $1/^{35}T_1$ is seen at
a temperature of $\sim$~14~K, labeled by a $T^*$ as shown in Fig.~\ref{invt1chp}.
Since the value of the $1/^{35}T_1$ already becomes too small below 100 K, this peaked behavior is
unlikely to be an intrinsic effect. Given that $T^*$ is close to the $T_N$ of the low-pressure phase (Fig.~\ref{invt1chp}),
it is possible that a residual low-pressure phase extends to high pressures with a very small volume fraction,
and affects the $1/^{35}T_1$ of the high-pressure phase by a proximity effect.

vi) {\it Transition to a magnetically disordered state.}
From all above discussions, we think that our data give compelling evidence that there is no magnetic ordering
at high pressures. As a result, a magnetically disordered phase is established, and labeled
in the high-pressure regime in the phase diagram in Fig.~\ref{pd}(a).

In particular, a magnetic valence-bond-crystal state may be formed,
in which the lattice symmetry is (A) either spontaneously
broken~\cite{Sachdev_NP_2008,Pujari_PRL_111087203_2013}
or (B) explicitly broken under pressure. The large spin gap revealed
in the $1/^{35}T_1$ indicates that ground sate may belong to
case (B), where the interactions explicitly break the $C_3$
symmetry. If one of the three bonds is fairly larger than the other two,
the ground state form disconnected dimers and the energy gap
is of the order of the exchange interactions.

In $\alpha$-RuCl$_3$, a pressure-induced dimerized state with strongly enhanced interactions
is possible for the following reasons. First, local lattice dimerization may be favored under pressure.
A recent LDA calculation shows that the on-site Coulomb repulsion $U$ plays an essential role for local lattice structure of
$\alpha$-RuCl$_3$~\cite{Kim_PhysRevB.93.155143}. A zigzag pattern is stabilized with a large $U/t$, whereas lattice dimerization is
formed at small $U/t$. With applying pressure, a reduction of $U/t$ is expected in principle, which may drive toward a dimerized
lattice structure. Second, the dimerized local exchanges can be greatly enhanced under lattice dimerization.
Finally, if the pressure is not completely isotropic, spatially anisotropic interactions may further be enhanced.

In literature, structural dimers were reported frequently to assist local singlets, such as
these observed in TlCuCl$_3$~\cite{Ruegg_prl_2008} and SrCu$_2$(BO$_3$)$_2$~\cite{Kageyama_prl_1999}.
Therefore, we think the valence-bond-crystal ground state is a candidate for
the high-pressure phase of $\alpha$-RuCl$_3$, due to interplay among local structures and
spin-lattice couplings.

\section{Summary and acknowledgment \label{summary}}

To summarize, our magnetization and NMR studies on $\alpha$-RuCl$_3$ reveal a non-monotonic change of $T_N$ at low pressures,
and a phase separation at intermediate pressures below 1.05 GPa.
Above 1.05~GPa, a new magnetic ground state is stabilized, which has significantly reduced static susceptibility
and low-energy spin fluctuations. We discussed possible scenarios to account for these observations.
Among them, a magnetically disordered state, in particular a valence-bond-crystal state
is proposed as a candidate for the high-pressure phase, originating from pressure-enhanced
spatially anisotropic interactions. Our results provide $\alpha$-RuCl$_3$ as an interesting system for sensitive tuning
of magnetic properties with the interplay among the spin and the lattice degrees of freedom.

The authors acknowledge the discussions with Profs. Youwen Long, Rong Yu, Liling Sun,
Qianghua Wang, Shunli Yu, and Wei Ji. This work was supported by the
National Science Foundation of China (Grant
Nos.~11374364, 11374143, 11574392, and 11674157), by the Ministry of Science and
Technology of China (Grant No.~2016YFA0300504), and by the Fundamental
Research Funds for the Central Universities and the Research Funds of Renmin
University of China (Grant No.~14XNLF08).

\end{document}